\documentclass[pdflatex,sn-mathphys-ay]{sn-jnl}

\usepackage{graphicx}%
\usepackage{multirow}%
\usepackage{amsmath,amssymb,amsfonts}%
\usepackage{amsthm}%
\usepackage{mathrsfs}%
\usepackage[title]{appendix}%
\usepackage{xcolor}%
\usepackage{textcomp}%
\usepackage{manyfoot}%
\usepackage{booktabs}%
\usepackage{algorithm}%
\usepackage{algorithmicx}%
\usepackage{algpseudocode}%
\usepackage{listings}%

\usepackage{orcidlink}

\theoremstyle{thmstyleone}%
\theoremstyle{thmstyletwo}%

\theoremstyle{thmstylethree}%

\begin{document}

\title[The Building Blocks of Classical Nonparametric Two-Sample Testing Procedures: Statistically Equivalent Blocks]{The Building Blocks of Classical Nonparametric Two-Sample Testing Procedures: Statistically Equivalent Blocks}

\author*{Chase Holcombe\,\orcidlink{0009-0008-8395-5780}}
\email{holcombe@southalabama.edu}

\affil*{\orgdiv{Department of Mathematics and Statistics}, \orgname{University of South Alabama}, \orgaddress{\city{Mobile}, \state{Alabama}, \country{USA}}}

\abstract{Statistically equivalent blocks are not frequently considered in the context of nonparametric two-sample hypothesis testing. Despite the limited exposure, this paper shows that a number of classical nonparametric hypothesis tests can be derived on the basis of statistically equivalent blocks and their frequencies. Far from being a moot historical point, this allows for a more unified approach in considering the many two-sample nonparametric tests based on ranks, signs, placements, order statistics, and runs. Perhaps more importantly, this approach also allows for the easy extension of many univariate nonparametric tests into arbitrarily high dimensions that retain all null properties regardless of dimensionality and are invariant to the scaling of the observations. These generalizations do not require depth functions or the explicit use of spatial signs or ranks and may be of use in various areas such as life-testing and quality control. In the manuscript, an overview of statistically equivalent blocks and tests based on these blocks are provided. This is followed by reformulations of some popular univariate tests and generalizations to higher dimensions. A brief simulation study and comments comparing the proposed methods to existing testing procedures are offered along with some conclusions.}


\maketitle
\section{Introduction}
\label{intro}

Ranks, signs, placements, order statistics, runs: these are words that are commonly associated with classical nonparametric or distribution-free testing procedures. On the other hand, the phrase "statistically equivalent blocks" is not so frequently uttered when motivating, deriving, or teaching nonparametric tests. The discussion of these statistically equivalent blocks, hereafter se-blocks, is generally reserved for the construction of distribution-free tolerance intervals \citep{krishnamoorthy2009statistical}. Using se-blocks to construct tolerance regions was originally proposed by \cite{wilks1942statistical} and extended to higher dimensionality by \cite{wald1943extension}, though the term "statistically equivalent block" was not coined until Tukey provided a discussion of these methods \citep{tukey1947non}. This tolerance interval approach continues to receive interest in a variety of fields, such as medicine and quality control \citep{young2020nonparametric,liu2024distribution,holcombe2024distribution}. However, se-blocks are infrequently mentioned in the hypothesis testing context. Certainly, some hypothesis testing procedures have been developed with se-blocks in mind, such as the tests proposed by \cite{wilks1961combinatorial}, \cite{AndersonSome}, and \cite{matthews1996nonparametric}, but many of these tests are not well known. In this paper, the author hopes to show that far from being irrelevant to nonparametric hypothesis testing, many classical two-sample testing procedures can be derived on the basis of se-block frequencies (hereafter block frequencies), which not only provides unity in the presentation of these procedures, but also allows for the straightforward generalization to arbitrarily high dimension without changing null properties.

To begin, we first consider the definition of an se-block and of a block frequency. In general, let $Y_1,Y_2,...,Y_n$ be an \textit{i.i.d.} from a continuous distribution with dimension $p$ and CDF $G(-)$. Using these random variables, the real $p$-dimensional space can be partitioned into $n+1$ se-blocks, which we denote $B_1,B_2,...,B_{n+1}$. Given another set of \textit{i.i.d.} values, $X_1,X_2,...,X_m$, from a continuous distribution with CDF $F(-)$, the block frequency for the $i$-th se-block, denoted $R_i$ is defined as the number of $X$ values in the se-block, $R_i = \sum_{j=1}^m I(X_j \in B_i) $. It is, perhaps, easiest to begin with $p=1$ and then generalize to higher dimensions. In the univariate case, the partitioned areas are immediately given by the space along the real number line between the observations, less than the minimum value, and greater than the maximum value. Denoting $Y_{(i)}$ as the $i$-th order statistic from the $Y$ sample, we may define each se-block in the univariate case as,

\begin{equation}
\label{univariate_block_def}
\begin{aligned}
B_1 &= (-\infty,Y_{(1)}], \\
B_2 &= (Y_{(1)},Y_{(2)}], \\
..., \\
B_n &= (Y_{(n-1)},Y_{(n)}], \text{ and} \\
B_{n+1} &= (Y_{(n)},\infty).
\end{aligned}
\end{equation}

Given these se-blocks in the univariate case, the block frequencies would be defined such that $R_i = \sum_{j=1}^m I(Y_{(i-1)} < X_j \leq Y_{(i)})$ for $i \in \{ 2,3,...,n \}$ and similarly for $i = 1,n+1$. When the $X$ and $Y$ values come from identical populations ($F(x) = G(x)$ for all $x$), the probability contents of each se-block or coverages, denoted $Q_1,Q_2,...,Q_{n+1}$, are distributed jointly as $(Q_1,...,Q_{n+1}) \sim Dirichlet(\boldsymbol{1})$. This results in a marginal distribution of $Q_i \sim Beta(1,n)$ for each se-block coverage and a marginal distribution of $\sum_{i=1}^k Q_i \sim Beta(k,n+1-k)$ for the combined coverage of a set of se-blocks \citep[p.~237--243]{WilksNonparametric}. The $Dirichlet(\boldsymbol{1})$ distribution is flat over the $n$ dimensional simplex, illustrating that the coverage values are uniformly distributed over this support. While this result is interesting, the remarkable feature of se-blocks is that these distributional properties continue to hold in arbitrarily high dimensions. The challenge is that it is not immediately clear how to divide the real $p$-variate space when generalizing to $p>1$. One way to generalize se-block to the multivariate setting begins by mapping the $p$-variate observations, $\boldsymbol{Y}_1,...,\boldsymbol{Y}_n$, to the real number line using some real valued function $\phi_1(-)$. Denote these values as $Y^*_i = \phi_1(\boldsymbol{Y}_i)$ for $i=1,2,...,n$. The $p$-dimensional area that $\phi_1(-)$ maps to the area less than the minimum $Y^*$ (or greater than the maximum $Y^*$) may then be partitioned as an se-block, $B_1$. Particularly, using the minimum, the se-block is defined as the set $B_1=\{ \boldsymbol{Y} \in \mathbb{R}^p ; \phi_1(\boldsymbol{Y}) < Y^*_{(1:n)} \}$. Excluding the minimum (or maximum) $Y^*$ used and the already partitioned area this process may be repeated with another function, $\phi_2(-)$, mapping the remaining $n-1$ values to the real number line. So one may set $Y^*_i = \phi_2(\boldsymbol{Y}_i)$ for $i=1,2,...,n-1$ and partition another se-block such that $B_2 = \{ \boldsymbol{Y} \in \mathbb{R}^p ; \phi_2(\boldsymbol{Y}) < Y^*_{(1:n-1)} ,\boldsymbol{Y}\notin B_1)$, using the minimum. This process of partitioning and excluding $Y^*$ values is repeated until the final value, $\phi_n(\boldsymbol{Y})$, partitions the remaining area into the final two se-blocks, resulting in $n+1$ se-blocks. The method described does not exhaustively define all ways of partitioning se-blocks, but it suffices for the present purposes. Indeed, this is a very brief summary of the method described by \cite{tukey1947non}, who provides a proof. To the same result, sometimes this process of partitioning is described in terms of "cutting functions" as in \cite{matthews1996nonparametric}. In this manuscript, we restrict our focus to functions that map the $p$-variate observations onto one of the $p$ coordinate axes since this results in se-block construction that is invariant to scaling or units.

\begin{figure}[ht]
    \centering
    \includegraphics[width=\linewidth]{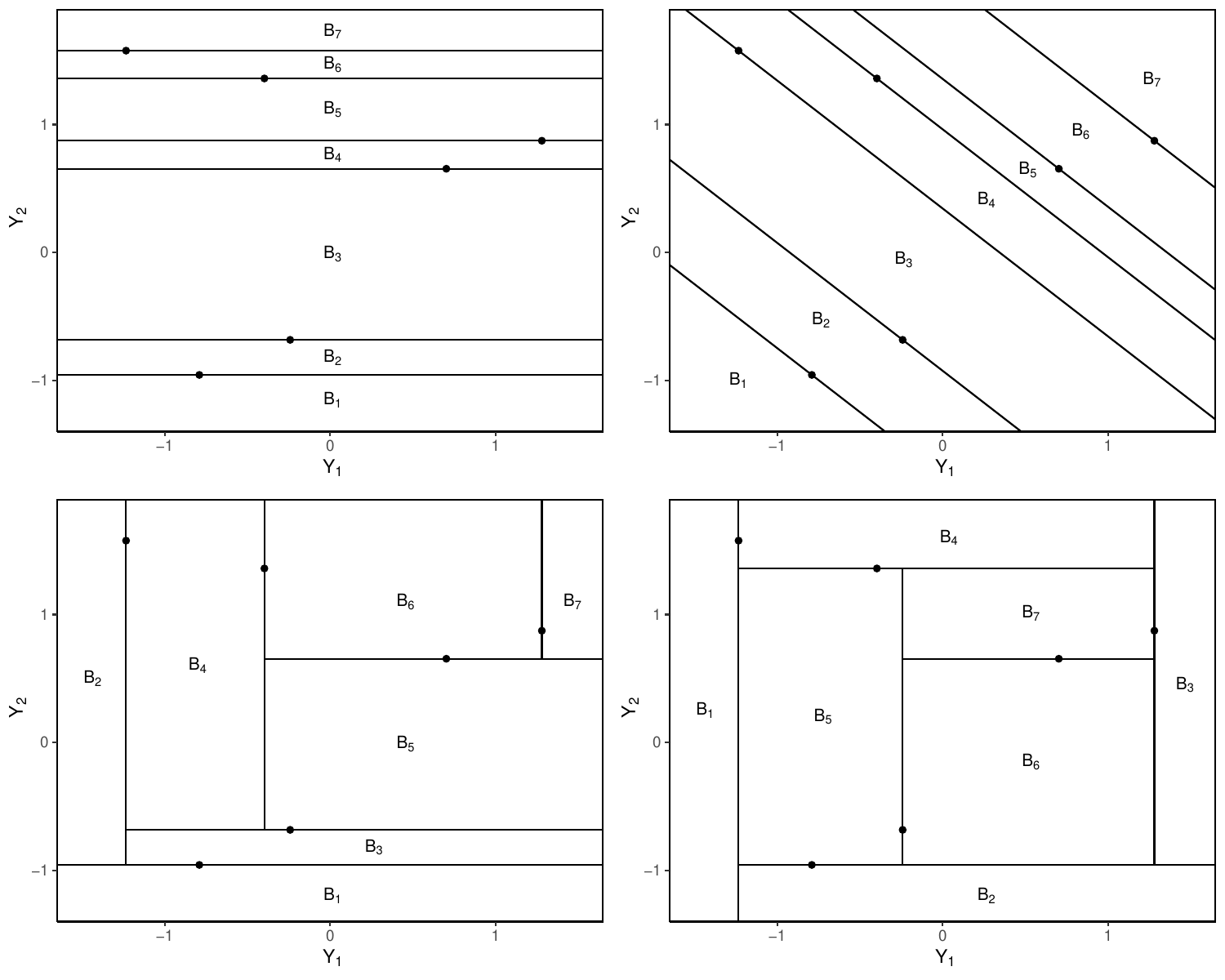}
    \caption{Four different ways to partition se-blocks using the example dataset found in Table \ref{tab:Data_toy}}
    \label{fig:block_construction}
\end{figure}

As a brief example to illustrate how se-blocks can be partitioned, consider the dataset in Table \ref{tab:Data_toy} and four example ways that se-blocks can be constructed using these observations in Figure \ref{fig:block_construction}. In the top left plot, the function $\phi_i((Y_1,Y_2))=Y_2$ is used for each $i$, which projects the observation onto the $Y_2$ axis. The minimum value is used to partition se-blocks at each step. A similar method is used in the top left plot with the minimums used to partition the se-blocks, but with the function $\phi_i((Y_1,Y_2))=Y_1 + Y_2$ for each $i$. This effectively projects the observations onto the line $Y_2 = Y_1$ and uses the ordering along this line. It is easy to see why this would not produce se-blocks invariant to scaling. On the bottom left, the function $\phi_i((Y_1,Y_2))=Y_2$ is used for odd $i$ and the function $\phi_i((Y_1,Y_2))=Y_1$ is used for even $i$, which alternates between projecting to the $Y_2$ axis (odd $i$) and $Y_1$ axis (even $i$). The minimum value is again used to partition se-blocks. Finally, on the bottom right, this alternating is also used, but with $\phi_i((Y_1,Y_2))=Y_2$ for odd $i$ and $\phi_i((Y_1,Y_2))=Y_2$ for even $i$. Here, the minimum is used to partition se-blocks $B_1$, $B_2$, and $B_5$ while the maximum is used to partition se-blocks $B_3$ and $B_4$. While there are uncountably many ways to partition the bivariate space into se-blocks using the example data in Table \ref{tab:Data_toy}, this provides a brief demonstration that illustrates the repeated method of ordering, partitioning, and excluding to construct se-blocks. As one final word of caution, we note that the order and way that the se-blocks will be partitioned, including the functions $\phi_i(-)$, should be decided before considering the data. Choosing a method to partition se-blocks that appear to be favorable \textit{after} viewing the data may lead to a departure from the theoretical distribution of the se-block coverages. This does not mean, however, that the method for partitioning se-blocks cannot be carefully chosen before viewing the data to achieve some desirable properties. In fact, we consider this point later in the manuscript.

\begin{table}[h]
    \centering
    \begin{tabular}{c|c c c c c c}
        $Y_1$ & 1.28 & -0.79 & 0.70 & -1.23 & -0.24 & -0.40 \\
        $Y_2$ & 0.87 & -0.96 & 0.65 & 1.58 & -0.68 & 1.36
    \end{tabular}
    \caption{An example dataset of six values from the standard bivariate normal distribution}
    \label{tab:Data_toy}
\end{table}

Now that we have considered the partitioning of se-blocks, we return to the subject of two-sample hypothesis testing procedures. The remainder of the manuscript is organized as follows. In Section \ref{exist_block_test}, we review some existing tests based on se-blocks, most of which are not well known. Then, in Section \ref{existing}, several existing two-sample univariate hypothesis tests that are not traditionally formulated using se-blocks are discussed and reformulated so to be based on se-blocks and block frequencies. Following this, in Section \ref{extend_to_multi}, several popular univariate nonparametric tests are extended to the multivariate setting. The extensions, based on se-blocks, are compared with existing multivariate nonparametric tests, such as those based on spatial signs and ranks. Finally, some conclusions are offered.

\section{Existing Tests Based on Statistically Equivalent Blocks}
\label{exist_block_test}

In this section, we consider some existing testing procedures that have been formulated based on se-blocks to test the general two-sample hypothesis, that both samples originate from the same population. Though outside the scope of this section, we briefly note that some hypothesis tests using se-blocks have been developed in the one-sample setting for goodness-of-fit. Before the term "statistically equivalent block" had been coined, \cite{fisher1929tests} proposed a one-sample test that some believe to have anticipated the concept \citep{AndersonSome,matthews1996nonparametric}. Similarly, \cite{david1950two} proposed an "empty cell" test derived via combinatorial properties that bears close resemblance to the "empty block" test, which we discuss in this section. More explicitly, the one-sample se-block tests have since been taken up by \cite{alam1993multivariate} and \cite{alam1995multivariate}. 

Perhaps the first two-sample test constructed on the basis of the distribution of block frequencies was proposed by \cite{dixon1940criterion}. Following the notation in Section \ref{intro}, this test statistic may be written as a function of block frequencies as

\begin{equation}
\label{Dixon_test}
C^2 = \sum_{i=1}^{n+1} (\frac{1}{n+1} - \frac{R_i}{m})^2.
\end{equation}

We must carefully avoid anachronism, though, since this test was proposed in the univariate case before se-blocks were explicitly proposed and refined by \cite{wald1943extension} and \cite{tukey1947non}. \cite{dixon1940criterion} proposed his test in the univariate context without using the term "block" and instead referenced the "elements ... on the same line ... divided into $n+1$ groups by the first sample." We note this test in this section since it is fundamentally a test on block frequencies and may be applied in arbitrarily $p$ dimensions with no necessary adjustment. Second, it is important background to the development of the other tests discussed in this section. \cite{mathisen1943method} proposed a similar test, specifically considering the frequencies of $X$ observations captured in each quartile form by the $Y$ observations. The fundamental idea behind these tests is that in the univariate case, one would expect the number of $X$ observations to be spread roughly evenly into the intervals between adjacent $Y$ values. More generally, one would not expect high deviations in block frequencies given the null hypothesis of identical distributions. Indeed, the $C^2$ statistic takes more extreme values if certain se-blocks contain heavier concentrations of $X$ values. In the context of these tests, \cite{blum1957consistency} proposed and studied several tests based on block frequencies. Like \cite{dixon1940criterion}, the term "block" was not mentioned and the tests were designed for the univariate setting. However, these tests may be taken to the multivariate setting with no necessary adjustments and Wilks specifically mentioned these tests in his treatment of two-sample nonparametric hypothesis tests based on se-blocks \citep[p.~450--451]{WilksNonparametric}. \cite{blum1957consistency} introduced the important concept of block frequency \textit{counts}. For $i = 0,1,...,m$, let $S_i$ be the random variable taking the \textit{count} of se-blocks containing $i$ values from the $X$ sample.

Following these developments, \cite{wilks1961combinatorial} proposed what is likely the first two-sample testing procedure to have been motivated by and explicitly defined using se-blocks and their frequencies. This test relies on block count frequencies, originally referred to by Wilks as \textit{cell} count frequencies to follow \cite{david1950two}, but quickly changed to \textit{block} count frequencies in his treatment less than two years later \cite[p.~444--445]{WilksNonparametric}. The test is based on the first $k$ block frequency counts for an arbitrary $k$. According to \cite{wilks1961combinatorial}, the test statistic "is essentially
that of the classical Pearson chi-square associated with the first $k + 1$ cell frequency counts," added to a term in order to produce the asymptotic distribution. As with the classical Pearson chi-square test, the goal of this test statistic is to capture deviation from the null distribution, in this case of the first $k$ block frequency counts. Soon after, a slightly less powerful but easier to construct testing procedure with an exact distribution for the test statistic was introduced on the basis of $S_0$, the number of empty se-blocks \citep[p.~446--452]{WilksNonparametric}. Indeed, $S_0$ is used directly as the test statistic. The distribution is given as in Equation (\ref{empty_block_mass}), which was originally derived on the basis of the joint distribution of $S_0,S_1,...,S_{m}$. However, we note that this distribution may also be derived on the basis of combinatorial properties.

As \cite{wilks1961combinatorial} pointed out, each vector of block frequencies $(r_1,...,r_{n+1})$ is equally likely with $\binom{m+n}{n}$ unique possible vectors, so $P(R_1=r_1,...,R_{n+1}=r_{n+1}) = \frac{1}{\binom{m+n}{n}}$. Thus, the probability of exactly $s_0$ empty se-blocks is equal to the number of unique arrangement producing $s_0$ empty se-blocks divided by $\binom{m+n}{n}$. Now out of the $n+1$ se-blocks there are $\binom{n+1}{s_0}$ ways to arrange the empty se-blocks. This leaves $n+1-s_0$ nonempty se-block. The number of ways to arrange the $m$ observations from the $X$ sample in these nonempty se-blocks so that all se-blocks contain at least a single $X$ is exactly $\binom{m-1}{n-s_0}$. Multiplying these terms yields the number of possible vectors of block frequencies such that $s_0$ se-blocks are empty such that,

\begin{equation}
\begin{gathered}
\label{empty_block_mass}
P(S_0 = s_0) = \frac{\binom{n+1}{s_0}\binom{m-1}{n-s_0}}{\binom{m+n}{n}}, \\ s_0 = \max\{0,n+1-m\},...,n-1,n.
\end{gathered}
\end{equation}

The empty block test statistic is easy to compute. The test may be constructed so that the null hypothesis of identical distributions is rejected whenever $S_0$ is large. This follows the same intuition driving the $C^2$ test from \cite{dixon1940criterion}. Given that the $X$ and $Y$ samples come from identical populations, one would expect the block frequencies to be roughly equal, with relatively few empty or highly concentrated se-blocks. This test, given the continuity assumption, has been shown to be consistent against all alternatives in the univariate case and, given "certain conditions" on the cumulative distribution functions, in the general $p$-variate case \cite[p.~447--452]{WilksNonparametric}. These factors make the test attractive, but as is the case with very general tests, the power is relatively low under many alternatives.

\cite{AndersonSome} followed soon after, proposing several hypothesis tests based on se-blocks for use in the multivariate case as well as a related two-sample classification procedure. one-sample goodness-of-fit tests were considered as well as two-sample hypothesis tests for equality of distributions. Though brief in the description of the two-sample testing procedure, Anderson provided several considerations which may be used in constructing a hypothesis test. Perhaps the most important point is that the "procedure for determining blocks" provides a natural ranking for the observations. Using the present notation, Anderson makes the point that by creating se-blocks $B_1,...,B_{n+1}$ with the $Y$ sample, the $X$ and $Y$ sets are naturally ranked such that for any $i$, $rank(Y_i|Y_1,...,Y_n)=k$ if $Y_i$ is the value partitioning se-blocks $B_k$ and $B_{k+1}$. Likewise, the $X$ observations may be ordered and ranked given the magnitude of $i$ for $X \in B_i$. Together, an $X$ observation is ordered and ranked above some $Y_i$ in the joint sample if and only if for $X \in B_k$, we have $rank(Y_i|Y_1,...,Y_n) < k$. As Anderson points, out, the assignment of ranks among the $X$ values in a given se-block $B_i$ "is immaterial," since these ties may be randomly broken. Ties are generally only a concern for rank based tests when the ties occur between values in different samples. Once the ranks have been assigned, any number of two sample rank tests may be used. Beyond this, Anderson does not provide much guidance, suggesting that the Wilcoxon-Mann-Whitney test could be used. In terms of appropriately partitioning the se-blocks, an illustration is given in which bivariate area is partitioned "in a spiral fashion," much like the bottom right illustration given here in Figure \ref{fig:block_construction}. Beyond this, in regard to the problem of partitioning se-blocks (or, equivalently, the choice of cutting functions) Anderson rhetorically asks, "Given a class of alternatives, what is the optimal choice?" His answer is that, "In general it is difficult to define the class of alternatives in a way that is suitable for formulating optimality." He does, however, mention the desirable properties of consistency and invariance, noting that for rank tests, "the probability of rejection [of the hypothesis of identical distribution] will approach 1 if the univariate test is consistent."

Since this, se-blocks have not received much attention in the two-sample hypothesis testing setting with the notable exception of \cite{matthews1996nonparametric}. Matthews and Taylor summarized existing one-sample and two-sample testing procedures based on se-blocks before considering a broader class of cutting functions, denoted as proximity-based cutting functions or PBCFs. In the language of \cite{tukey1947non}, this approach basically involves setting each function $\phi_i(-)$ to the same function where this mapping allows for se-blocks to be constructed based on proximity to the $Y$ observations. In this context, "proximity" is defined using Euclidean distance, which results in se-blocks that not invariant to scaling. Further, these se-blocks do not look block-like at all, as "the blocks are the areas bounded by level sets." Once the se-blocks have been partitioned, \cite{matthews1996nonparametric} propose following the general guidance of \cite{AndersonSome}, employing a rank test with the newly imposed ordering dictating the ranks. About their own procedure, they note, "The concept of PBCF holds promise for the analysis of multivariate data. Additional research is clearly in order. The power of the procedure has not been investigated." We do not pursue this method of partitioning se-blocks further here since we have restricted our focus to procedures that are unaffected by units or scaling, but it appears that this seemingly forgotten method may hold interesting results and a comprehensive power analysis remains to be done.

As is evident by this brief overview, the literature surrounding two-sample hypothesis tests formulated and defined using se-blocks is minuscule compared to the broader classical nonparametric hypothesis testing literature. Furthermore, there has seemingly been little interest in se-blocks in the two-sample testing context since the 1960s. However, we now turn our attention to some well-known two-sample hypothesis tests to show that many popular procedures can be formulated and defined with se-blocks and block frequencies, which makes the extension to the multivariate setting straightforward.

\section{Some Popular Univariate Two-Sample Nonparametric Hypothesis Tests}
\label{existing}

\subsection{Precedence and Maximal Precedence Tests}
\label{prec_tests}

We first take up the precedence test and related precedence-type tests. A precedence-type test was originally proposed in the "exceedance" context by \cite{gumbel1950distribution}. The exceedance statistic takes the number of values in one-sample that exceed a specified order statistic from the other sample. Given that the two samples come from the identical continuous populations, the number of exceedances follows what has since come to be known as the negative hypergeometric distribution. This exceedance test is linearly related to the more popular precedence test, which was soon taken up by various individuals, especially in the context of life-testing situations \citep{nelson1963tables,balakrishnan2006precedence}. \cite{gumbel1950distribution} referred to their work on the exceedance test "as a
specialization and a generalization of the work done by Wilks," in reference to \cite{wilks1942statistical}. As is stated in the introduction, this paper originally proposed the univariate nonparametric tolerance interval that was generalized by \cite{wald1943extension} and discussed by \cite{tukey1947non}. So it is no surprise that there exists a tight link between precedence-type tests and se-blocks. Without loss of generality, we focus here on the precedence test, counting the precedences among the $X$ sample. Given that the precedence statistic takes the number of $X_1,...,X_m$ that precede the $j$-th order statistic from $Y_1,...,Y_n$, the mass function of the precedence statistic, $W_j$, is given by

\begin{equation}
\label{prec_mass}
P(W_j=w) = \frac{\binom{w+j-1}{w}{\binom{m-w+n-j}{m - w}}}{\binom{m+n}{n}}, w = 0,1,...,m.
\end{equation}

It was established early that the null distribution of precedence-type tests can be derived on the basis of combinatorial properties \citep{epstein1954tables} or by convolution \citep{nelson1963tables}. Using the later method, one is required to use the distribution of $Y_{(j)}$, which is known to be $Beta(j,n-j+1)$, as in \citep[p.~37--39]{gibbons2014nonparametric}. The marginal distributions of the order statistics themselves, however, may be derived based on the fact that the $n$ order statistics partition the real number line into $n+1$ se-blocks and, thus, the the area less than some $Y_{(j)}$ is the combined area of $j$ se-blocks. Thus, we consider a generalized formulation for the precedence test.

The test statistic for this generalized precedence test, $T_j$, takes the number of $X$ values that fall into some subset of $j$ se-blocks from $B_1,B_2,...B_{n+1}$ for $1 \leq j \leq n$. The coverage for this set of se-blocks is then $\sum_{i=1}^j Q_i$. Since $Q_1,...,Q_{n+1} \sim Dir(\boldsymbol{\alpha})$, it follows that $\sum_{i=1}^j Q_i \sim Beta(j,n+1 - j)$. The number of precedences will then follow the binomial distribution with $m$ trials and success probability $\sum_{i=1}^j Q_i$. So, the mass function for $T_j$ is derived by the following:

\begin{equation}
\begin{aligned}
\label{gen_prec_mass}
P(T_j = t) &= \int_0^1 \binom{m}{t} q^t (1-q)^{m-t} * \frac{q^{j-1}(1-u)^{n-j}}{B(j,n+1-j)} dq \\
&= \frac{\binom{t+j-1}{t} \binom{m-t+n-j}{m - t}}{\binom{m+n}{n}}, t = 0,1,...,m.
\end{aligned}
\end{equation}

The mass function for $T_j$ in Equation (\ref{gen_prec_mass}) is equivalent to the mass function for $W_j$ in Equation (\ref{prec_mass}). Thus the precedence test is a special case of this generalized test, when the first $j$ se-blocks, $B_1,B_2,...,B_j$, are chosen, as defined as in Equation (\ref{univariate_block_def}). Similarly, the exceedance test is a special case when the upper-most $j$ se-blocks are selected.

In addition to the simplest forms of the precedence and exceedance tests, variants with improved power under some alternatives have been introduced. One popular test is the maximal precedence test, proposed by \cite{balakrishnan2000precedence} and \cite{balakrishnan2001general}. Proposed in the life-testing context, the associated test statistic takes "the maximum number of failures occurring from the $X$-sample before the first, between the first and the second,..., between the $(r-1)$th and $r$th failures from the $Y$-sample," for some $0 < r \leq n$ \citep[p.~61--62]{balakrishnan2006precedence}. More generally, this test takes the maximum number of values from the $X$ sample that fall into a single interval out of the $j$ intervals formed by the first $j$ order statistics among the $Y$ sample, rather than the total number of $X$ values in this area, as with the traditional precedence test. The connection with se-blocks is immediately clear as this test statistic is simply taking the value of the maximum block frequency among $j$ se-blocks for some $0 < j \leq n$. The joint distribution of any $j$ block frequencies, denoted $R_1^*,R_2^*,...,R_j^*$, is known to have the mass function given by the following \citep[p.~442--443]{WilksNonparametric}:

\begin{equation}
\begin{gathered}
\label{joint_block_mass}
P(R_1^* = r_1^*, R_2^* = r_2^*, ..., R_j^* = r_j^*) = \frac{\binom{m + n - \sum_{i=1}^j r_i - j}{n - j}}{\binom{m+n}{n}}, \\ r_i = 0,1,...,m; \sum_{i=1}^j r_i \leq m.
\end{gathered}
\end{equation}

The mass function for the maximum may be derived by first noting that the joint distribution for all block frequencies, once again given by Wilks, is $[\binom{m+n}{n}]^{-1}$, illustrating that all possible values for the length $n+1$ vector of block frequencies are equally likely \citep[p.~442--443] {WilksNonparametric}. So the mass function may be derived as the quotient of the sum of ways in which the maximum value equals some $r_{max}$ and the total number of arrangements for the frequencies. This is taken up in more detail in Section \ref{ext_prec}, but it here suffices to say that these details agree with the treatment given by \cite{balakrishnan2006precedence} in deriving the distribution of the maximal precedence statistic. The connection is that the maximal precedence statistic is a special case of a so-called maximal block frequency statistic, when the first $j$ se-blocks from Equation (\ref{univariate_block_def}) are selected.

In this section, some popular precedence-type tests have been related to se-blocks and block frequencies. We return to these tests in Section \ref{extend_to_multi}, where we consider the generalization to higher dimensions. First, we continue to another popular univariate two-sample nonparametric test based on runs.

\subsection{The Wald–Wolfowitz Runs Test}
\label{suns_test}

The Wald-Wolfowitz Runs Test (hereafter runs test) is a general two-sample procedure designed to test the null hypothesis of equality of two continuous univariate distributions. It was originally proposed by \cite{wald1940test} and later served as the basis for the related and similarly named runs test for randomness from \cite{wald1943random}. Here, we focus on the two-sample testing procedure. It is well known that the runs test is very general, which has the advantage in that it is consistent for all types of differences between the two populations \citep[p.~452--454] {WilksNonparametric}, but struggles from weak "performance against specific alternatives," \cite[p.~234]{gibbons2014nonparametric}. Nonetheless, the test remains popular as an easy-to-use nonparametric testing procedure. The test statistic, denoted $U$, takes the number of runs from the combined sample, $X_1,...,X_m, Y_1,...,Y_n$, once it has been put in ascending order. A run is defined as a sequence of $X$ or $Y$ values in the ordered combined sample. Consider the example in Table \ref{tab:cauchy_ex_identical}. Here, we have nine observations $X_1,...,X_4,X_5,Y_1,...,Y_4$ from the same $Cauchy(0,1)$ distribution, which have been arranged in ascending order with the "current" run labeled. In this example, $U=6$.

\begin{table}[h]
    \centering
    \begin{tabular}{c|c c c c c c c c c c}
        Value: & -4.62 & -1.56 & -0.36 & -0.21 & 0.00 & 0.13 & 0.27 & 0.75 & 3.32  \\
        Sample & X & X & Y & X & Y & X & X & Y & Y \\
        Run: & 1 & 1 & 2 & 3 & 4 & 5 & 5 & 6 & 6 \\
    \end{tabular}
    \caption{An example for the runs test with observations from identical populations}
    \label{tab:cauchy_ex_identical}
\end{table}

Intuitively, the observations in Table \ref{tab:cauchy_ex_identical} appear to be well mixed, which is consistent with the null hypothesis of identical distributions. Indeed, the runs test when applied to these data does not result in a rejection at any conventional significance level. Now consider the observations presented in Table \ref{tab:cauchy_ex_dif}. These $X$ and $Y$ observations were drawn from the $Cauchy(0,1)$ distribution and the $Cauchy(10,1)$ distribution respectively. In this example $U=2$. Since the data are not mixed well, it would seem that there is strong evidence against the hypothesis of equality of distributions. Indeed, the runs test rejects the null for these data at the $\alpha = 0.05$ level.

\begin{table}[h]
    \centering
    \begin{tabular}{c|c c c c c c c c c c}
        Value: & -1.89 & 1.77 & 2.25 & 1.23 & -0.94 & 9.53 & 11.43 & 5.91 & 9.70 \\
        Sample & X & X & X & X & X & Y & Y & Y & Y \\
        Run: & 1 & 1 & 1 & 1 & 1 & 2 & 2 & 2 & 2 \\
    \end{tabular}
    \caption{An example for the runs test with observations from different populations}
    \label{tab:cauchy_ex_dif}
\end{table}

The theoretical details for the runs test, including asymptotic properties, have been derived. The mass function, identical to the mass function for the distribution of groups considered earlier by \cite{stevens1939distribution}, is a piecewise function given by

\begin{equation}
\begin{gathered}
\label{runs_mass}
P(U=u) = \left\{
\begin{array}{ll}
      2 \binom{m-1}{u/2 - 1} \binom{n - 1}{u/2 - 1} [\binom{m + n}{n}]^{-1} & \text{for $u$ even} \\
      ( \binom{m-1}{(u - 1)/2} \binom{n - 1}{(u - 3)/2} + \binom{m-1}{(u - 3)/2} \binom{n - 1}{(u - 1)/2} ) [\binom{m + n}{n}]^{-1} & \text{for $u$ odd,} \\
\end{array} 
\right. \\
u = 2,3,...,\min \{2n+1,2m+1,m+n \}.
\end{gathered}
\end{equation}

The mass function bears relation to that of the empty block test statistic, but it is certainly not identical. At this point, the connection between runs and block frequencies may not be immediately clear. However, we now turn out attention to examining the relationship between the runs test and the empty block test more carefully. Upon closer investigation, the connection between these two testing procedures is tighter than is initially evident. Despite considering the runs test immediately after the empty block test, Wilks makes only the comment, "Note that the random variables $u$ and $s_0$ are strongly negatively correlated," where $u$ and $s_0$ are the runs test and empty block test statistics, respectively \cite[p.~452]{WilksNonparametric}. Indeed, this is so since a larger number of runs necessitates fewer empty se-blocks. To illustrate this point, reconsider the $Y$ sample from Table \ref{tab:cauchy_ex_identical}. By constructing se-blocks as in Equation (\ref{univariate_block_def}), we have the following se-blocks:

\begin{equation}
\label{example_runs_block}
\begin{aligned}
B_1 &= (-\infty,-0.36], \\
B_2 &= (-0.36,0.00], \\
B_3 &= (0.00,0.75], \\
B_4 &= (0.75,3.32], \text{ and} \\
B_5 &= (3.32, \infty).
\end{aligned}
\end{equation}

Now, if an $X$ sample of size $m\geq5$ were drawn, the number of runs would be some value $2 \leq U \leq 9$. Note that $U=9$ if and only if there is an $X$ observation in each se-block (i.e. all block frequencies are nonzero.). Similarly, $U = 8$ is only possible if and only if either $B_5$ is empty or $B_1$ is empty. Consider now lower numbers of runs. We have $U=2$ only if all $X$ values are in $B_1$ or all $X$ values are in $B_5$. Likewise, $U=3$ if and only if either all $X$ values are in one se-block out of $B_2,B_3,B_4$ or if $X$ values are in both $B_1$ and $B_5$, but not in $B_2,B_3,B_4$.

To formalize the relationship between empty se-blocks and runs, we make one further distinction that likely has little relevance outside of the univariate case. For se-blocks formed from $Y_1,...,Y_{n+1}$, denote se-blocks $B_1$ and $B_{n+1}$ as \textit{exterior se-blocks} and denote all other se-blocks as \textit{interior se-blocks}. In general, for $U=u$, we have the following cases:

\begin{enumerate}
    \item If $u$ is even, then $\frac{2n-u}{2}$ interior se-blocks are empty and 1 exterior se-block is empty.
    \item If $u$ is odd, then one of the following cases applies: 
    \begin{itemize}
        \item $\frac{2n-u-1}{2}$ interior se-blocks are empty and both exterior se-blocks are empty.
        \item $\frac{2n-u+1}{2}$ interior se-blocks are empty and no exterior se-blocks are empty.
    \end{itemize}
\end{enumerate}

Thus, a test based on runs is equivalent to a test based on the number of empty interior and exterior se-blocks. This is not identical to the empty block test discussed in Section \ref{exist_block_test}. The test statistic does not take the number of empty se-blocks, $S_0$, but rather is a function of the number of empty interior se-blocks and empty exterior se-blocks, denoted $S_{0,IN}$ and $S_{0,EX}$, respectively. We may confirm this by considering the joint distribution of the number of interior and exterior se-blocks. The joint distribution of $S_{0,IN}$ and $S_{0,EX}$ may be derived by a combinatorial argument, much like the distribution of $S_0$ in Section \ref{exist_block_test}. The number of arrangements that guarantee some $S_{0,IN}=s_{0,IN}$ and $S_{0,EX}=s_{0,EX}$ is equal to the product of the number of ways that the empty exterior se-blocks can be arranged, the number of ways that the empty interior se-blocks can be arranged, and the number of ways that the $m$ observations from the $X$ sample can be arranged in the $n+1-s_{0,IN}-s_{0,EX}$ nonempty se-blocks such that each of these se-blocks contains at least one $X$ observation. These pieces are $\binom{2}{s_{0,EX}}$, $\binom{n-1}{s_{0,IN}}$, and $\binom{m-1}{n-s_{0,IN}-s_{0,EX}}$, respectively. Thus, the joint mass function for $S_{0,IN}$ and $S_{0,EX}$ is this product divided by the number of equally likely arrangements for the $n+1$ block frequencies, as in the following:

\begin{equation}
\begin{gathered}
\label{int_ext_empty_block_mass}
P(S_{0,IN}=s_{0,IN},S_{0,EX}=s_{0,EX}) = \frac{\binom{2}{s_{0,EX}} \binom{n-1}{s_{0,IN}} \binom{m-1}{n-s_{0,IN}-s_{0,EX}}}{\binom{m+n}{n}}, \\
s_{0,IN} = 0, 1,..., n - 1, \\ s_{0,EX} = 0,1,2, \\ \max \{0,n+1-m \} \leq s_{0,IN} + s_{0,EX} \leq n.
\end{gathered}
\end{equation}

Using Equations (\ref{int_ext_empty_block_mass}) and (\ref{runs_mass}), the following is easily verified:

\begin{enumerate}
\item $P(U=u) = P(S_{0,IN}=\frac{2n-u}{2},S_{0,EX}=1)$ when $u$ is even
\item $P(U=u) = P(S_{0,IN}=\frac{2n-u-1}{2},S_{0,EX}=2 \cup S_{0,IN}=\frac{2n-u+1}{2},S_{0,EX}=0)$ when $u$ is odd
\end{enumerate}

Thus, it has now been shown that the runs test is fundamentally a test based on se-blocks and block frequencies. Since the concept of internal and external se-blocks likely do not have relevance beyond the univariate case, we do not consider multivariate generalizations of this test in Section \ref{extend_to_multi} but instead note that the existing empty block test, as discussed in Section \ref{exist_block_test}, serves as an approximate multivariate generalization of this test. We now turn our attention to two-sample testing procedures based on ranks and placements.

\subsection{The Mann-Whitney, Wilcoxon Rank-Sum, and Related Tests}
\label{rank_tests}

In this section, we consider two types of tests that are closely related, namely tests based on placement and tests based on ranks. Though similar, placements are distinct from ranks. In the univariate case, the placement of some $X$ among a set of values $Y_1,...,Y_n$ is defined as $\sum_{i=1}^n I(Y_i \leq X)$, where $I(-)$ is the indicator function. We first consider tests based on placements since the connection between placements, precedence values, and se-blocks is quite clear. Indeed, the placement of $X$ among the set of $Y$ values is equivalent to the number of $Y$ values preceding or equal to $X$ and we have shown that precedence values are closely related to se-blocks in Section \ref{prec_tests}. In this context, several testing procedures have been proposed that rely on placements and precedences \citep{orban1982class,fligner1976some}. Perhaps the most popular test based on placements is the Mann-Whitney test \citep{mann1947test}. Using the present notation, this test is based on the statistic defined by
\begin{equation}
\label{mw_u_def}
U = \sum_{j=1}^m \sum_{i=1}^n I(Y_i \leq X_j).
\end{equation}
It was proposed to test the null hypothesis of stochastic equality against the alternative of first order stochastic dominance with two samples from continuous distributions. Though $U$ is typically regarded as a function of the placements of the $X$ values among the $Y$ sample, \cite{mann1947test} originally proposed the test using the "precedence" terminology.

Intuitively, given the null, one would expect the $X$ values to be roughly evenly spread across the $Y$ values so that the sum of the placements is neither abnormally high (indicating that the $X$ stochastically dominates the $Y$) nor abnormally low (indicating that the $Y$ stochastically dominates the $X$). The distribution for $U$, and thus the critical values, may be found via a recursive relationship, though the mass function does not have an analytically available form without including the recursive terms. Also, as \cite{gibbons2014nonparametric} point out, even for $m,n$ as small as 6, the asymptotic $N(\frac{mn}{2},\frac{mn(m+n+1)}{12})$ distribution is "reasonably accurate" for approximating the exact distribution.

Returning to the definition of the placement, we note that when the underlying population is continuous, a placement value of $k$ for some $X$ among the $Y$ is equivalent to $X \in (Y_{(k)}, Y_{(k+1)}]$ for $k = 1,...,n-1$. Similarly, placements of $0$ and $n$ imply $X \in (-\infty,Y_{(1)}]$ and $X \in (Y_{n},\infty)$, respectively. Thus, these placements may be represented in terms of \textit{block placements}. So $U$ may be equivalently represented by

\begin{equation}
\begin{aligned}
\label{mw_u_block}
U &= \sum_{j=1}^m \sum_{i=0}^n i * I(X_j \in B_{i+1}) \\
&= \sum_{i=0}^n i * R_{i+1}.
\end{aligned}
\end{equation}

Thus, the Mann-Whitney test may be defined using se-blocks. Indeed, the motivation for using se-blocks follows the intuition driving most of the nonparametric procedures discussed in this paper, but it especially resembles the motivation behind the precedence and maximal precedence procedures discussed in Section \ref{prec_tests} as the test statistic can be written as a function of the block frequencies, $R_1,...,R_{n+1}$. Under the null hypothesis of identical distributions, one would not expect large differences in the block frequencies. The connection between se-blocks and placements that we point out is not unprecedented. One popular nonparametric textbook, relating ranks, placements, and se-blocks, notes that the placement is "simply equal to" the sum of the block frequencies \citep[p.~65]{gibbons2014nonparametric}. On the same page, it is similarly noted that ranks and placements are closely related, with the result that, using the present notation, the rank of $X_{(k)}$ in the combined sample of the $X$ and $Y$ values is $\sum_{i=1}^n I(Y_i \leq X_{(k)}) + k$. Thus, the placement of an $X$ among the $Y$ sample is linearly related to the rank of the $X$ value in the combined sample. This is a well known result and connects the Mann-Whitney test with the Wilcoxon rank-sum test.

The Wilcoxon rank-sum test, was originally proposed by Frank Wilcoxon in a very brief paper, along with the similarly named one-sample signed-rank test \citep{wilcoxon1945individual}, before the Mann-Whitney test was proposed. Though the test is often presented in the context of the location problem, the rank-sum test was originally proposed without an explicit alternative, following Fisher's significance testing paradigm. Where $\boldsymbol{Z} = (Z_1,...,Z_{m+n})$ is the indicator vector such that $Z_i$ denotes whether the $i$-th ranked value in the combined sample belongs to the $X$ sample ($Z_i = 1$) or to the $Y$ sample ($Z_i = 0$), the rank-sum statistic is defined as

\begin{equation}
\label{RS_W}
W = \sum_{i=1}^{m+n} i * Z_i.
\end{equation}

Since we have now shown that ranks are closely related to placements and placements are closely related to se-blocks, it would seem that the rank-sum test statistic can be easily written in terms of se-blocks. One difficulty, however, is that the order statistic $X_{(k)}$, from the $m$ $X$ observations for some $k$, is not uniquely defined outside of the univariate case. One way to overcome this challenge would be to directly define the indicator vector $\boldsymbol{Z}$ in terms of se-blocks. This may be done by defining the elements of $\boldsymbol{Z}$ with the following rule: Each element $Z_i$ such that $i = \sum_{k=1}^j R_k + j$ for $j \in \{1,2,...,n\}$ is equal to $0$. All other elements of $\boldsymbol{Z}$ are equal to 1. This is equivalently given by
\begin{equation}
\label{Z_general}
Z_i = 
\begin{cases} 
0 & i \in \{ R_1 + 1, \sum_{k=1}^2R_k + 2,...,\sum_{k=1}^nR_k + n \}, \\
1 & \text{otherwise}. 
\end{cases}
\end{equation}
Importantly, this reduces to the usual definition of $\boldsymbol{Z}$ in the univariate case when $B_1,...,B_{n+1}$ are defined as in Equation (\ref{univariate_block_def}).

We may think about the indicator vector $\boldsymbol{Z}$ corresponding to a sort of ranking procedure where the $X$ values in an se-block $B_i$ are ranked higher than the $X$ values in se-blocks $B_1,...,B_{i-1}$ and lower than the $X$ values in se-blocks $B_{i+1},...,B_{n+1}$. Similarly, the $X$ values in $B_i$ are ranked higher than exactly $i-1$ values from the $Y$ sample. If the se-blocks are partitioned in ascending order $B_1,B_2,...$, then we may think of each $X \in B_i$ being ranked higher than the $Y$ values that partitioned $B_1,...,B_{i-1}$. This has the attractive property in that for the purposes of linear rank tests, this generalization results in the same indicator vector $\boldsymbol{Z}$ as previously defined in terms of ranks. Using the generalized indicator vector $\boldsymbol{Z}$, the test statistics for other popular location tests such as the Terry-Hoeffding Normal Scores \citep{terry1952some,hoeffding1951optimum} and van der Waerden \citep{waerden1953order} tests as well as those for scale tests such as the Mood \citep{mood1954asymptotic}, Klotz Normal Scores \citep{klotz1962nonparametric}, and Siegel-Tukey \citep{siegel1960nonparametric} tests may be defined in terms of block frequencies. For the sake of simplicity, however, we may also define the rank-sum test statistic in relation to the Mann-Whitney test statistic. \cite{mann1947test} pointed out that $U$ is linearly related to $W$ by a shift of size $\frac{m(m+1)}{2}$ (or $-mn - \frac{m(m+1)}{2}$ depending on whether the $X$ or $Y$ ranks are summed). So the rank-sum test statistic may be written in terms of block frequencies as in

\begin{equation}
\label{RS_block}
W = \frac{m(m+1)}{2} + \sum_{i=0}^n i * R_{i+1}.
\end{equation}

We have now shown that the Mann-Whitney test as well as all linear rank tests may be defined in terms of se-blocks. Indeed, as we turn out attention to extending these tests in the general multivariate setting, the only item to be changed is the se-block construction, which does not change the null distribution of the test statistic.

\section{Extending Popular Univariate Tests to the Multivariate Case}
\label{extend_to_multi}

\subsection{SE-Block Partitioning Methods}
\label{block_part_methods}

We now turn our attention to extending some of these popular univariate tests to the multivariate setting using the properties of se-blocks. Here, denote the two samples from $p$-variate populations as $\boldsymbol{X}_1,...,\boldsymbol{X}_m$ and $\boldsymbol{Y}_1,...,\boldsymbol{Y}_n$, where $\boldsymbol{X}_i = (X_{i,1},...X_{i,p})$ and $\boldsymbol{Y}_i = (Y_{i,1},...Y_{i,p})$, with the order of the components arbitrarily selected. Since the previously discussed univariate tests have now been defined on the basis of se-blocks, they may be easily extended given that the se-blocks are defined in different ways, suitable to the test and the class of alternatives considered. Thus, we first consider a few ways to partition se-blocks in $p$ dimensions that we use in this section, though it is noted that there are theoretically many more ways to partition $n+1$ se-blocks with the $\boldsymbol{Y}$ sample, uncountably infinitely many ways if procedures dependent on scaling or units are considered. To motivate this discussion, consider, once again, the se-block constructions in Figure \ref{fig:block_construction}, discussed in Section \ref{intro}. One could partition se-blocks as in the bottom left of Figure \ref{fig:block_construction}, by partitioning the minimum (maximum) among the first component, $Y_{1,1},Y_{2,1},...,Y_{n,1}$, and subsetting, then partitioning among the minimum (maximum) among the second component, $Y_{1,2},Y_{2,2},...,Y_{n,2}$, and subsetting, and continuing until the minimum (maximum) is partitioned among the $p$-th component, $Y_{1,p},Y_{2,p},...,Y_{n,p}$. Then, the process is repeated, beginning with the first and continuing up to the $p$-th component. Whether the se-blocks are partitioned in ascending or descending order for a given component may be determined arbitrarily or based on the specific class of alternatives considered. More specifically, using the ascending order, this involves using functions $\phi_1(\boldsymbol{Y_i}) = Y_{i,1}$, $\phi_2(\boldsymbol{Y_i}) = Y_{i,2}$, ... $\phi_p(\boldsymbol{Y_i}) = Y_{i,p}$, $\phi_{p+1}(\boldsymbol{Y_i}) = Y_{i,1}$, ... for any $i$. A slight adjustment would be use to use the stair-step method, but to reverse the component ordering on every other run, so that we have $\phi_1(\boldsymbol{Y}_i) = Y_{i,1}$, ..., $\phi_p(\boldsymbol{Y}_i) = Y_{i,p}$, $\phi_{p+1}(\boldsymbol{Y}_i) = Y_{i,p}$, ..., $\phi_{2p}(\boldsymbol{Y}_i) = Y_{i,1}$, $\phi_{2p+1}(\boldsymbol{Y}_i) = Y_{i,1}$, ... for any $i$. We may refer to this type of partitioning method as the \textit{stair-step} method.

An alternative method for partitioning se-blocks may be to partition from the most extreme values in an inward direction, as in the bottom right of Figure \ref{fig:block_construction}. Following this example closely, we may partition in the same way as the stair-step method, but alternate between using the maximum and minimum each time that we project onto one of the coordinate axes. So the $\phi(-)$ functions are the same as in the stair-step method with the only difference being that the partitioning values (maximum or minimum) alternates each time some $\phi(\boldsymbol{Y}_i)=Y_{i,j}$, for some component value $j$. One slight adjustment would be to iteratively partition both the minimum and maximum with a given projection onto a coordinate axis before continuing to the next component, following \cite{AndersonSome} in his rank-sum example. In either case, we may call this the \textit{spiral} method since the se-blocks are partitioned in a circular fashion from the outside towards the center of the data. These are just a two general types of methods that may be used to partition se-blocks, which we utilize in the remainder of this section.

\subsection{Tests Based on Block Frequency Summaries}
\label{ext_prec}

As we have illustrated in Section \ref{prec_tests}, the precedence and maximal precedence tests are fundamentally tests based on block frequencies. We first turn our attention to the precedence test. Given some hypotheses, one could construct blocks using either the star step or the spiral method. Then, for some $j$, with $T_j$ defined as in Section \ref{prec_tests}, $T_j$ follows the negative hypergeometric distribution as in Equation \ref{gen_prec_mass}. The following questions remain: what is the correct null and alternative hypotheses, what is the appropriate choice for $j$, which method for constructing blocks in the $p$-dimensional space should be utilized, and does it make sense to conduct a one-tailed or two-tailed test. It is difficult to answer these questions separately since they are interconnected, so we consider two cases.

Consider first the multivariate lifetime setting. In the context of reliability, the problem of multivariate life-testing has been considered by various individuals \citep{miller1977note,ghurye1987some}. However, the distribution-free multivariate lift testing question has not been as carefully considered. While it may seem reasonable to assume independence in some settings in order to use univariate testing procedures with a multiple testing correction, \cite{gupta2010} point out that, "in many reliability situations, it is more realistic to assume some form of dependence among components." It is precisely in this setting that a generalized form of the precedence test may be useful. With life-testing in mind, the null hypothesis of identical populations may be tested against an alternative of stochastic dominance. This is commonly employed for the precedence test in the univariate setting and is a more general alternative containing "the location shift alternative and the Lehmann alternative" as special cases \cite[p.~32]{balakrishnan2006precedence}. The concept of stochastic dominance in the multivariate setting is more complex since it it not uniquely defined, but it would seem that a component-wise dominance, like that considered by \cite{o1991multivariate}, would be an appropriate alternative. In the univariate setting, \cite{bowker1944note} showed that the control median test (which is in fact a special case of the precedence test) is not consistent against the entire class of stochastically ordered alternatives. However, the precedence test has been shown to be consistent against the narrower location-shift alternative \citep{chakraborti1996precedence}. For the generalized form of the precedence test considered here, it is also easy to construct examples, regardless of the se-block construction method, where the test is not consistent against a component-wise stochastically ordered alternative. However, it is also easy to construct small classes of alternatives against which the multivariate test is consistent. It remains an interesting area of future research to identify large classes of alternatives against which this test is consistent.

Correctly identifying a relevant subset of alternatives may also help answer the remaining questions. The stair-step method for partitioning se-blocks appears to be correct for this setting, since the value of the test statistic, $T_j$, would roughly correspond with more or less precedences component-wise. So whether the test is two-sided or one-sided depends on whether the stochastic dominance alternative is strictly ordered. However, we note that this test construction has the potential to "mask" evidence against the null hypothesis if, for example, the marginal distribution for one component of the $\boldsymbol{X}$ sample dominates the corresponding component in the $\boldsymbol{Y}$ sample while the marginal distribution for a different component among the $\boldsymbol{X}$ sample is dominated by the corresponding component in the $\boldsymbol{Y}$ sample. The "masking" effect has been noted in the univariate case \citep[p.~61]{balakrishnan2006precedence}, and in both the univariate and multivariate case the choice of $j$ can impact which subset of stochastically ordered alternatives are "masked" in this testing procedure. Also, $j$ may be chosen to accommodate some right-censoring level among the $\boldsymbol{Y}$ sample.

Moving from the life-testing setting to the more general case, it may make more sense to construct the se-blocks using the spiral procedure. This design, while not immune to the "masking" effect, could avoid "masking" a location shift among one or more components. In this case $T_j$ can be thought of as corresponding to the number of $\boldsymbol{X}$ values outside of the hyper-rectangular region formed by se-blocks $B_{j+1},B_{j+2},...,B_{n+1}$, which we may denote $C$. In the very general case of testing the null hypothesis of identical populations against the alternative of nonidentical populations, it may make sense to construct a two-sided test. However, it is once again easy to construct alternatives against which this test is not consistent. So it is perhaps more interesting to consider a test against the location-shift alternative. More $\boldsymbol{X}$ values falling outside of $C$ constitute evidence in favor of the alternative. So it would be appropriate to construct an upper-tail rejection region for $T_j$, using Equation \ref{gen_prec_mass}. The value $j$ should be chosen to ensure the existence of a rejection region given the sample sizes and may be selected based off of power under specified alternatives. However, as a general rule of thumb, it seems reasonable to construct the two regions (inside and outside the hyper-rectangle) using roughly an equal number of blocks, which implies $j \approx \frac{n+1}{2}$. It would seem that under the class of location-shift alternatives, this test is consistent under the conditions noted by \cite{chakraborti1996precedence}. However, this remains to be rigorously shown.

Now, turning to the maximal precedence test, we note that this procedure can be similarly extended. In the multivariate lifetime setting, it would seem appropriate to test the null hypothesis of identical distributions against the alternative of stochastic dominance, as with the generalized precedence test. By partitioning se-blocks using the stair-step method, the test would be sensitive to component-wise stochastic dominance in either direction. Following \cite{balakrishnan2001general}, the one-sided test is appropriate. The intuition is the same as in the empty block test. Under the null hypothesis of identical distributions, the $\boldsymbol{X}$ values are expected to be roughly equally spread across the $n+1$ se-blocks, with would lead to a relatively smaller value for the maximal precedence test statistic. However, under the alternative, one would expect the $\boldsymbol{X}$ values to be more heavily concentrated in some se-blocks than others, resulting in a large maximal precedence test statistic. The choice of $j$ may be chosen arbitrarily or, as is common in the life-testing setting, to accommodate right-censoring among the $\boldsymbol{Y}$ sample. In general, if no censoring occurs, it may also be reasonable to let $j=n+1$, meaning that the test statistic takes the maximum block frequency. In general, larger values for $j$ result in less information loss. Though \cite{balakrishnan2001general} only considered $j \leq n$, the null distribution and properties of the univariate test hold in this multivariate extension. Where $R_{(j)}^{MAX}$ is the maximum block frequency out of the first $j$ se-blocks, with serves as the test statistic, the mass function is given by the following:
\begin{equation}
\begin{gathered}
\label{max_prec_mass}
P(R_{(j)}^{MAX} = r) = \sum_{(r_1,...,r_j) \in S_j} \frac{\binom{m + n - \sum_{i=1}^j r_i - j}{n - j}}{\binom{m+n}{n}} * I(\max[r_1,...,r_j] = r), \\
S_j = \{ (r_1,...,r_j) | (r_1,...,r_j \in \mathbb{Z}^+) \cap \\ [(\sum_{i=1}^j r_i \leq m \cap j \leq n) \cup (\sum_{i=1}^j r_i = m \cap j = n + 1) ] \}.
\end{gathered}
\end{equation}
Note that $S_j$ is the support set containing vectors of achievable block frequencies, $(r_1,...,r_j)$.

Like the precedence test, this generalized maximal precedence test is also appropriate outside of the lifetime setting. As previously stated, if censoring is not present, it seems appropriate to utilize all of the available information and take the maximum block frequency in all $n+1$ se-blocks, which means setting $j=n+1$. Identifying appropriate classes of alternatives for this particular test remains an area of future research. Similarly, using the stair-step or spiral method for partitioning se-blocks could result in a different testing decision for a particular dataset, but it is not apparent that either method holds an advantage in terms of test properties under alternatives of a location shift or component-wise stochastic dominance. As a final note on this test, it is perhaps more informative to refer to this hypothesis testing procedure as a \textit{maximal block test}, illustrating the similarity to the empty block test.

In this section, we have extended the precedence and maximal precedence tests to the multivariate setting. These tests hold interesting potential particularly in the life-testing context due to the ability to easily handle right censoring among the $\boldsymbol{Y}$ sample. This is a practical issue that may occur in various cases, including the manufacturing and medical settings \cite[p.~31--32]{balakrishnan2006precedence}. Even outside of the lifetime setting, these tests are useful in that they are exactly distribution-free for all continuous distributions. When compared with the existing empty block test, these procedures also have the advantage in that they may easily handle imbalanced data. Consider the case of a large $\boldsymbol{X}$ sample and a small $\boldsymbol{Y}$ sample. Given that the $\boldsymbol{Y}$ sample is used to partition the se-blocks, it is very unlikely that many, or perhaps any, se-blocks would be empty. Similarly, if the labels are flipped and thus the larger sample is used to partition the se-blocks, the problem remains. Due to sparsity, it is very unlikely that many, if any, block frequencies are greater than one. Both the generalized precedence and maximal precedence (maximal block) tests are at an advantage in this situation. Particularly, if the smaller sample is labeled as the $\boldsymbol{Y}$ sample and used to partition the se-blocks, both tests are flexible and likely to provide a non-null rejection region at conventional significance levels even when $m >> n$.

\subsection{Tests Based on SE-Blockwise Ranks}
\label{ext_rank}

We have shown in Section \ref{exist_block_test} that \cite{AndersonSome}, in very general fashion, illustrated that the Wilcoxon Rank-Sum or Mann-Whitney tests can be extended to the multivariate setting. Indeed, we have further shown in Equation (\ref{Z_general}) that the indicator vector $\boldsymbol{Z}$ can be defined in terms of block frequencies, which provides a more general extension to all linear rank tests. Thus, given some block partition for the $\boldsymbol{Y}$ sample, the elements $Z_1,...,Z_{m+n}$ are easily calculated. This allows for an order to be imposed among the two samples, namely that $\boldsymbol{X}$ values in $R_i$ are ordered below $Y_i$ and above $Y_{i-1}$. Linear rank statistics may then be calculated in a straightforward fashion as a function of $\boldsymbol{Z}$. As in the univariate setting, the mean, variance, and covariance for the elements of $\boldsymbol{Z}$ are given by the following \cite[p.~277]{gibbons2014nonparametric}:

\begin{equation}
\begin{gathered}
\label{properties_z}
E[Z_i] = \frac{m}{m+n}, \\
Var[Z_i] = \frac{mn}{(m+n)^2}, \\
Cov[Z_i,Z_j] = \frac{-mn}{(m+n)^2(m+n-1)}, \\
1 \leq i , j \leq m+n ; i \neq j.
\end{gathered}
\end{equation}

The extension of the indicator vector $\boldsymbol{Z}$ to the multivariate setting allows for any two-sample distribution-free tests with a test statistic $T = \sum_{i=1}^{m+n} a_i Z_i$ to be extended to arbitrarily high dimension. As has been pointed out, this class of tests includes many popular two-sample location and scale tests, with various weights $a_i$ \cite[Chapter~7--9]{gibbons2014nonparametric}. With this extension, the null properties of each test remain unchanged regardless of the dimension.

Once $\boldsymbol{Z}$ is defined using se-blocks, the multivariate extension is straightforward. As in Section \ref{ext_prec}, the remaining challenge is in choosing an appropriate partitioning method when constructing the se-blocks and in identifying a relevant subset of alternatives for each specific testing procedure. For the partitioning procedure, the spiral method appears to be natural for an extension, since, for ascending values $a_i$, the resulting test would be sensitive to location or scale shifts. The stair-step method could also be desirable for settings such as life-testing. Due to the large number of linear rank tests with test statistic $T = \sum_{i=1}^{m+n} a_i Z_i$, it is not now appropriate to examine the properties of each test under various alternatives. However, to illustrate the testing procedure and the generalization of $\boldsymbol{Z}$, we turn to a simulated example.

Consider the data in Tables \ref{tab:null_data_Z} and \ref{tab:alt_data_Z}. The former is drawn from the standard bivariate normal distribution, as with the data in Tables \ref{tab:Data_toy}, and the latter is drawn from the same distribution with a shift of magnitude 1 in the negative direction along both coordinate axes.

\begin{table}[ht]
    \centering
    \begin{tabular}{c|c c c c c c c c}
        $X_1$ & -0.69 & -1.13 & -0.92 & 2.21 & 1.02 & -1.57 & 1.20 & 0.22 \\
        $X_2$ & -0.18 & 0.33 & -0.87 & 0.67 & -2.14 & -1.04 & -1.42 & 0.34
    \end{tabular}
    \caption{An example dataset of eight values from the standard bivariate normal distribution}
    \label{tab:null_data_Z}
\end{table}

\begin{table}[ht]
    \centering
    \begin{tabular}{c|c c c c c c c c}
        $X_1$ & -0.25 & -2.21 & 0.11 & -1.45 & 0.64 & 0.81 & -3.18 & -2.18 \\
        $X_2$ & -1.79 & -0.26 & -1.66 & -1.42 & -1.66 & -1.88 & -2.01 & -0.61
    \end{tabular}
    \caption{An example dataset of eight values from the bivariate normal distribution with mean vector $(-1,-1)^T$}
    \label{tab:alt_data_Z}
\end{table}

\begin{figure}[ht]
    \centering
    \includegraphics[width=\linewidth]{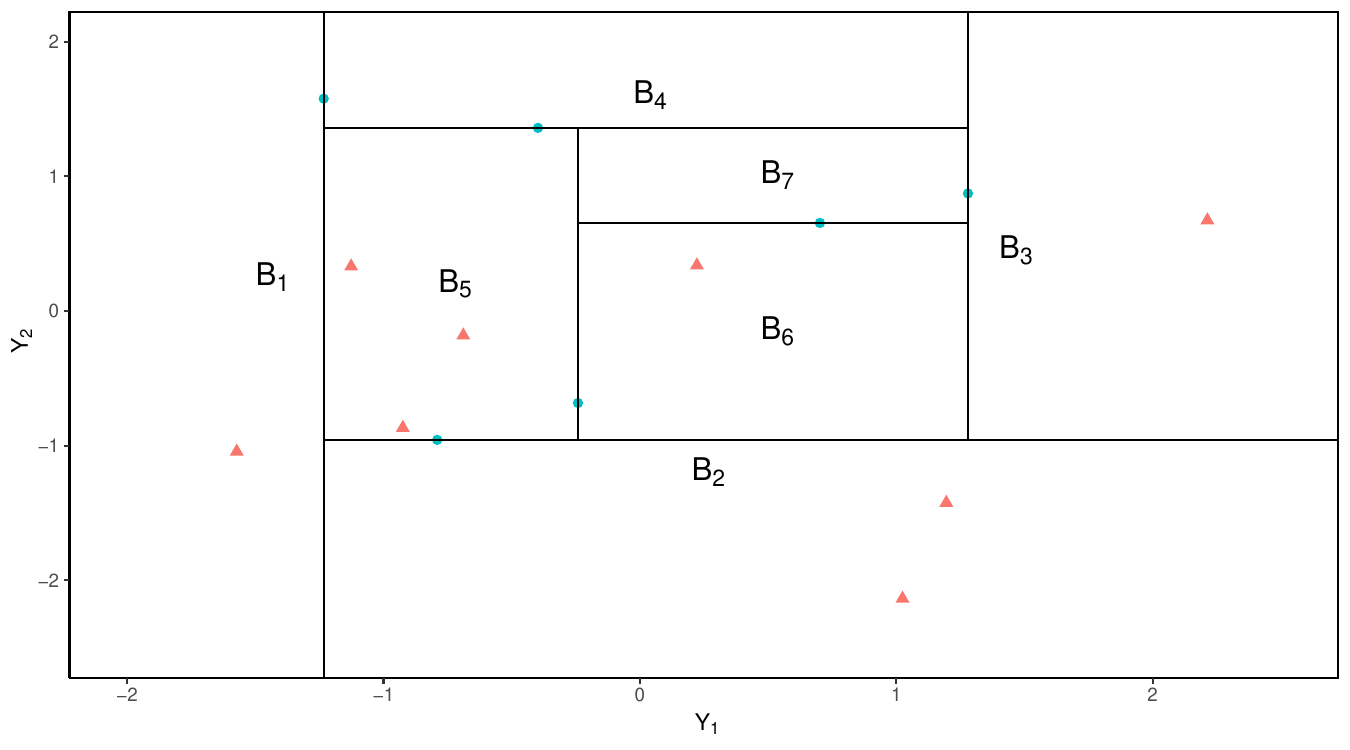}
    \caption{Scatter plot of observations from Tables \ref{tab:Data_toy} and \ref{tab:null_data_Z} with partitioned se-blocks}
    \label{fig:null_Z_examp}
\end{figure}

\begin{figure}[ht]
    \centering
    \includegraphics[width=\linewidth]{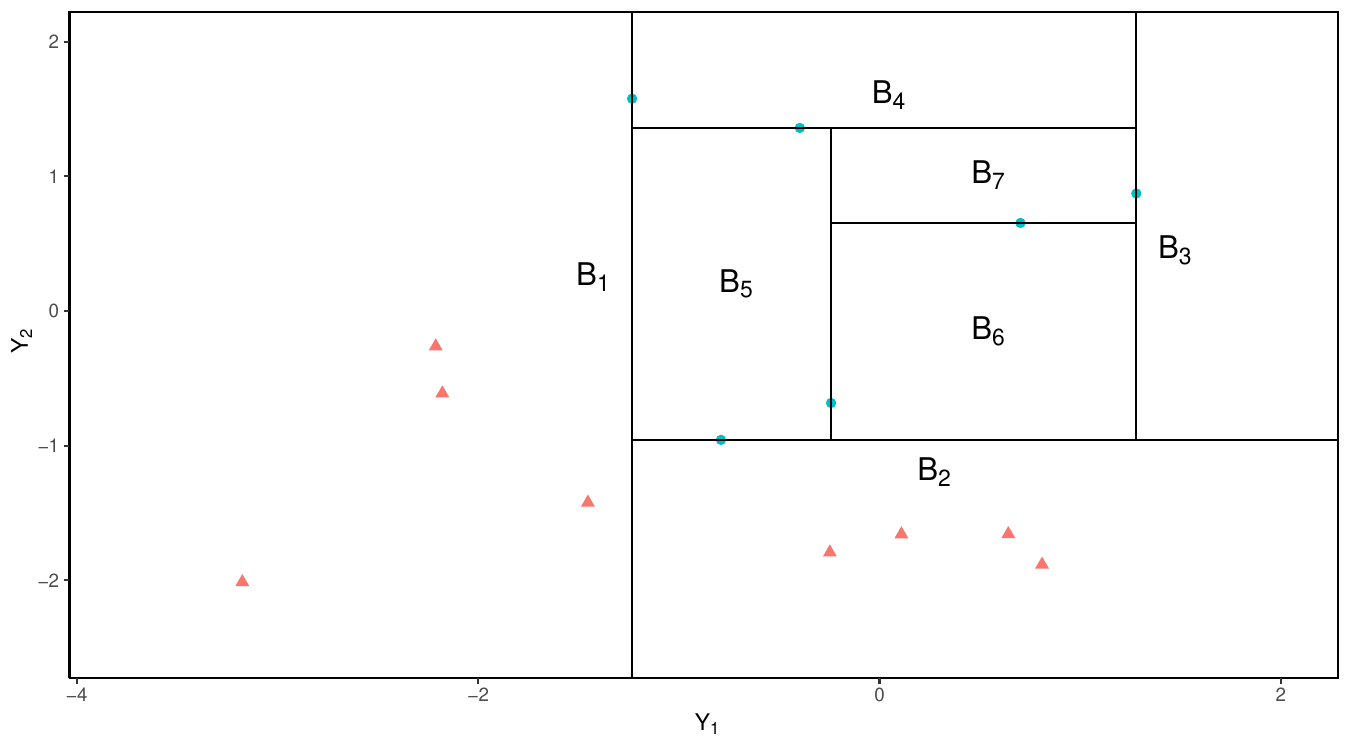}
    \caption{Scatter plot of observations from tables \ref{tab:Data_toy} and \ref{tab:alt_data_Z} with partitioned se-blocks}
    \label{fig:alt_Z_examp}
\end{figure}

To illustrate the linear rank tests in the multivariate setting, we set up hypothesis tests between the data from these datasets (the $\boldsymbol{X}$ sample) and the data from Table \ref{tab:Data_toy} (the $\boldsymbol{Y}$ sample). In the null and alternative cases, displayed in Figures \ref{fig:null_Z_examp} and \ref{fig:alt_Z_examp}, respectively, it may be observed that the block frequencies are as in Table \ref{tab:block_freqs_example}. From these block frequencies, the indicator vector can be calculated as

\begin{equation}
\label{Z_null}
\boldsymbol{Z}_{\text{null}} = (1,0,1,1,0,1,0,0,1,1,1,0,1,0),
\end{equation}
in the null case and as

\begin{equation}
\label{Z_alt}
\boldsymbol{Z}_{\text{alt}} = (1,1,1,1,0,1,1,1,1,0,0,0,0,0),
\end{equation}
 in the alternative case.

\begin{table}[ht]
    \centering
    \begin{tabular}{c|c c c c c c c}
        Block Frequency: & $R_1$ & $R_2$ & $R_3$ & $R_4$ & $R_5$ & $R_6$ & $R_7$ \\
        Null Example: & 1 & 2 & 1 & 0 & 3 & 1 & 0 \\
        Alternative Example: & 4 & 4 & 0 & 0 & 0 & 0 & 0 \\
    \end{tabular}
    \caption{The block frequencies $R_1,...,R_7$ corresponding to Figures \ref{fig:null_Z_examp} and \ref{fig:alt_Z_examp}}
    \label{tab:block_freqs_example}
\end{table}

With the values of the indicator vector $\boldsymbol{Z}$ specified, the linear rank tests can be performed. In Table \ref{tab:obs_sig}, p-values and values of the linear rank statistic corresponding to some popular two-sample linear rank tests are reported for both the null example and the alternative example.

\begin{table}[ht]
    \centering
    \begin{tabular}{c|c c c}
        Test & Wilcoxon & Terry-Hoeffding & van der Waerden \\
        $\sum_{i=1}^{m+n} a_i \boldsymbol{Z}_{\text{null}}$ & 57 & -0.9410 & -0.8012 \\
        Null p-value & 0.755 & 0.614 & 0.616 \\
        $\sum_{i=1}^{m+n} a_i \boldsymbol{Z}_{\text{alt}}$ & 40 & -4.474 & -4.0764 \\
        Alternative p-value & 0.008 & 0.005 & 0.011 \\
    \end{tabular}
    \caption{The observed significance level for several two-sample tests}
    \label{tab:obs_sig}
\end{table}

It may be observed from this illustration that the tests detect the shift in the alternative example at conventional significance level. This example should not be taken as indicating any relative performance between the tests in detecting a location shift, but is provided for illustrative purposes to show how $\boldsymbol{Z}$ is calculated and the tests are performed in the multivariate setting. Thus, we have demonstrated how linear rank tests may be generalized to the multivariate setting using se-blocks with the null properties of these tests remaining unchanged. We now turn our attention to a brief power study and some comments comparing tests based on se-blocks and existing nonparametric multivariate tests.

\subsection{Comparison with Existing Nonparametric Multivariate Tests}
\label{spatial_comp}

Given that so many univariate tests are formulated in terms of signs and ranks, it is no surprise that several multivariate testing procedures have been formulated in terms of \textit{spatial} signs and ranks. Many methods for assigning spatial signs or ranks have been proposed with this goal of extending popular one-sample and two-sample nonparametric tests to the multivariate setting. In particular, the testing procedure proposed by \cite{mottonen1995multivariate} imposes an ordering "using $L_1$ criterion functions," with the spatial sign function returning "the unit vector in the direction" of the observation \citep{mottonen1997efficiency}. The spatial rank function is defined similarly. \cite{oja2004multivariate} proposed a similar procedure, but instead utilized the $L_2$ criterion in defining the spatial sign function. The strength of these methods is that the spatial signs and ranks are fairly easy to compute using simple matrix operations and are either rotation invariant, as in \cite{mottonen1995multivariate}, or are affine invariant, as in \cite{oja2004multivariate}. However, none of these methods are exactly distribution-free and are instead \textit{asymptotically} distribution-free. In addition to continuity, each testing procedure also relies on the assumption of symmetry for convergence of the test statistic to some known distribution. In contrast with these methods, the proposed testing procedures based on se-blocks are \textit{exactly} distribution-free and require only the assumption of continuity. Fundamentally, tests based on spatial signs or ranks are only asymptotically distribution-free due to the properties of the spatial signs or ranks themselves.

More recently, \cite{chernozhukov2017monge} and \cite{hallin2021distribution} have proposed spatial ranks based on measure transportation. Building on this, \cite{deb2023multivariate} proposed multiple new multivariate hypothesis testing procedures based on the rank-energy statistic, including a general two-sample test for equality of distributions, which is \textit{exactly} distribution-free and does not require symmetry. Speaking about this method, the authors said, "As far as we are aware, this is the first attempt to systematically develop distribution-free multivariate tests that are consistent against all alternatives and are computationally feasible," while noting that the methods proposed by \cite{rosenbaum2005exact} and \cite{boeckel2018multivariate} also fit these categories, with the latter being similar to their proposed method. Perhaps overlooked, the older and perhaps simpler empty block test discussed in Section \ref{exist_block_test} is also exactly distribution-free, consistent against all alternatives, and computationally feasible given some mild assumptions. Several extensions similar to and building on the measure transportation based spatial ranks have since been proposed. The interested reader is referred to the discussion section from \cite{deb2023multivariate} for more details. The testing procedure proposed by \cite{rosenbaum2005exact} is quite different than methods based on spatial ranks and is easy to implement using the \texttt{crossmatch} package available on CRAN \citep{crossmatchpackage}. The general idea of this testing procedure is to pair the observations in the pooled using some distance-based criterion. Then, the number of "cross-match" pairs, which are pairs containing an observation from each sample, is taken to be the test statistic. The method is only invariant to rotation or changes in scale given some scale and rotation invariant distance metric, such as Mahalanobis distance.

Since we have shown that using se-blocks allows for distribution-free tests in the univariate case (whether based on signs, ranks, placements, or runs) to be easily extended to the multivariate case, we see that these generalized tests join a small group of \textit{exactly} distribution-free multivariate testing procedures. Nonetheless, given the existence of other \textit{exactly} distribution-free multivariate hypothesis testing procedures, the question arises as to whether the tests based on se-blocks maintain competitive power or not. We present a small power study in Tables \ref{tab:pow_loc}, \ref{tab:pow_scale}, and \ref{tab:pow_overcon} with sample size $m=n=200$, dimension $p=3$, significance level $\alpha=0.05$, and number of replicates $N=10,000$. Under various alternatives, simulated power values are reported for the rank-sum (RS), Terry-Hoeffding (TH), van der Waerden (VdW), generalized precedence (Prec.), maximal block (MB), and empty block (EB) tests using both the spiral (Sp) and stair-step (SS) constructions. The $\boldsymbol{X}$ and $\boldsymbol{Y}$ samples are randomly assigned as the partitioning sample and the columns are randomly permuted before the test is performed. Simulated power values for the rank-energy (RE) \citep{deb2023multivariate} and cross-match (CM) \citep{rosenbaum2005exact} tests are also reported for comparison. For all tests with discrete test statistics, randomization is used to guarantee the significance level $\alpha = 0.05$. Since normal theory methods have been shown to struggle against the so-called "mixture alternative" \citep{rosenblatt2018mixture}, several scenarios have been selected where only a portion of the population shifts in location or scale. In addition to the location and scale shifts, we have also included two scenarios where an over-concentrated region exists in one population, not unlike the perturbation alternatives discussed in \cite{sriperumbudur2010hilbert} and \cite{gretton2012kernel}. The six alternative scenarios are as listed below, with each scenario utilizing the covariance (for Normal) or scale (for Cauchy) matrix with elements $\Sigma_{ij}=0.35^{|i-j|}$:

\begin{flalign*}
    \text{Alternative 1: }& \boldsymbol{X}\sim Cauchy(\boldsymbol{0},\Sigma) \text{ and} && \\
    & \boldsymbol{Y}\sim 0.9 \cdot Cauchy(\boldsymbol{0},\Sigma) + 0.1 \cdot
    Cauchy((0,c,c)',\Sigma) && \\
    & \text{for }c=5,10,15; &&
\end{flalign*}
\begin{flalign*}
    \text{Alternative 2: }& \boldsymbol{X}\sim Normal(\boldsymbol{0},\Sigma) \text{ and} && \\
    & \boldsymbol{Y}\sim 0.9 \cdot Normal(\boldsymbol{0},\Sigma) + 0.1 \cdot
    Normal((0,c,c)',\Sigma) && \\
    & \text{for }c=2,2.25,2.5; &&
\end{flalign*}
\begin{flalign*}
    \text{Alternative 3: }& \boldsymbol{X}\sim Normal(\boldsymbol{0},\Sigma) \text{ and} && \\
    & \boldsymbol{Y}\sim Normal(\boldsymbol{0},c \cdot \Sigma) && \\
    &\text{for } c = 1.5,2,2.5; &&
\end{flalign*}
\begin{flalign*}
    \text{Alternative 4: }& \boldsymbol{X}\sim Normal(\boldsymbol{0},\Sigma) && \\
    & \boldsymbol{Y}\sim 0.9 \cdot Normal(\boldsymbol{0},\Sigma) + 0.1 \cdot
    Cauchy(\boldsymbol{0},c \cdot \Sigma) && \\
    & \text{for } c = 2,4,6; &&
\end{flalign*}
\begin{flalign*}
    \text{Alternative 5: }& \boldsymbol{X}\sim Cauchy(\boldsymbol{0},\Sigma) \text{ and} && \\
    & \boldsymbol{Y}\sim (1 - c) \cdot Cauchy(\boldsymbol{0},\Sigma) + c \cdot
    Uniform([0.45,0.55]^3) && \\
    & \text{for } c = 0.1,0.2,0.3; &&
\end{flalign*}
\begin{flalign*}
    \text{Alternative 6: }& \boldsymbol{X}\sim Normal(\boldsymbol{0},\Sigma) \text{ and} && \\
    & \boldsymbol{Y}\sim (1 - c) \cdot Normal(\boldsymbol{0},\Sigma) + c \cdot
    Uniform([0.45,0.55]^3) && \\
    & \text{for } c = 0.1,0.2,0.3. &&
\end{flalign*}

\begin{table}[ht]
    \centering
    \begin{tabular}{c|c c c | c c c}
        & & Alt. 1 & & & Alt. 2 & \\
        \hline
        & $c = 5$ & $c = 10$ & $c = 15$ & $c = 2$ & $c = 2.25$ & $c = 2.5$\\
        \hline
        RS (Sp) & 0.1849 & 0.2479 & 0.2851 & 0.1900 & 0.2273 & 0.2552 \\
        VdW (Sp) & 0.1450 & 0.2259 & 0.2630 & 0.2146 & 0.2531 & 0.2928 \\
        TH (Sp) & 0.1531 & 0.2124 & 0.2625 & 0.2170 & 0.2615 & 0.2995 \\
        Prec. (Sp) & 0.1635 & 0.1774 & 0.1953 & 0.1206 & 0.1461 & 0.1541 \\
        MB (Sp) & 0.1682 & \textbf{0.3604} & \textbf{0.4393} & 0.1183 & 0.1609 & 0.2175 \\
        EB (Sp) & 0.1028 & 0.1578 & 0.1942 & 0.0999 & 0.1135 & 0.1305 \\
        RS (SS) & 0.1853 & 0.2333 & 0.2508 & 0.2162 & 0.2463 & 0.2531 \\
        VdW (SS) & 0.1788 & 0.2540 & 0.2817 & 0.2717 & 0.2981 & 0.3224 \\
        TH (SS) & 0.1802 & 0.2446 & 0.2777 & 0.2774 & 0.2980 & 0.3280 \\
        Prec. (SS) & 0.1358 & 0.1341 & 0.1375 & 0.1356 & 0.1407 & 0.1394 \\
        MB (SS) & 0.1394 & 0.2591 & 0.3143 & 0.1097 & 0.1404 & 0.1851 \\
        EB (SS) & 0.0892 & 0.1312 & 0.1512 & 0.0894 & 0.0981 & 0.1098 \\
        RE & \textbf{0.2576} & 0.3016 & 0.3245 & \textbf{0.2830} & \textbf{0.3097} & \textbf{0.3300} \\
        CM & 0.1069 & 0.1334 & 0.1501 & 0.0999 & 0.1161 & 0.1257
    \end{tabular}
    \caption{The power of several tests under alternative scenarios 1 and 2}
    \label{tab:pow_loc}
\end{table}

\begin{table}[ht]
    \centering
    \begin{tabular}{c|c c c | c c c}
        & & Alt. 3 & & & Alt. 4 & \\
        \hline
        & $c = 1.5$ & $c = 2$ & $c = 2.5$ & $c = 2$ & $c = 4$ & $c = 6$\\
        \hline
        RS (Sp) & 0.9134 & 0.9998 & \textbf{1.0000} & 0.0884 & 0.1696 & 0.2221 \\
        VdW (Sp) & 0.9267 & \textbf{0.9999} & \textbf{1.0000} & 0.0906 & 0.2118 & 0.2908 \\
        TH (Sp) & \textbf{0.9301} & \textbf{0.9999} & \textbf{1.0000} & \textbf{0.0911} & \textbf{0.2164} & \textbf{0.2928} \\
        Prec. (Sp) & 0.7608 & 0.9962 & \textbf{1.0000} & 0.0717 & 0.1060 & 0.1297 \\
        MB (Sp) & 0.1295 & 0.3765 & 0.6284 & 0.0484 & 0.0568 & 0.0615 \\
        EB (Sp) & 0.1672 & 0.5421 & 0.8727 & 0.0528 & 0.0751 & 0.0921 \\
        RS (SS) & 0.1754 & 0.4048 & 0.6089 & 0.0580 & 0.0606 & 0.0677 \\
        VdW (SS) & 0.1763 & 0.4000 & 0.5764 & 0.0511 & 0.0629 & 0.0763 \\
        TH (SS) & 0.1765 & 0.3865 & 0.5736 & 0.0510 & 0.0680 & 0.0749 \\
        Prec. (SS) & 0.1447 & 0.3184 & 0.4875 & 0.0507 & 0.0524 & 0.0587 \\
        MB (SS) & 0.0981 & 0.2469 & 0.4321 & 0.0530 & 0.0515 & 0.0574 \\
        EB (SS) & 0.1109 & 0.2808 & 0.5274 & 0.0513 & 0.0685 & 0.0750 \\
        RE & 0.1314 & 0.5273 & 0.9283 & 0.0523 & 0.0561 & 0.0541 \\
        CM & 0.1491 & 0.4854 & 0.8242 & 0.0579 & 0.0676 & 0.0729
    \end{tabular}
    \caption{The power of several tests under alternative scenarios 3 and 4}
    \label{tab:pow_scale}
\end{table}

\begin{table}[ht]
    \centering
    \begin{tabular}{c|c c c|c c c}
        & & Alt. 5 & & & Alt. 6 & \\
        \hline
        & $c = 0.1$ & $c = 0.2$ & $c = 0.3$ & $c = 0.1$ & $c = 0.2$ & $c = 0.3$ \\
        \hline
        RS (Sp) & 0.2603 & \textbf{0.7557} & \textbf{0.9640} & 0.2255 & \textbf{0.7061} & \textbf{0.9404} \\
        VdW (Sp) & 0.2372 & 0.7111 & 0.9053 & 0.2041 & 0.6544 & 0.8693 \\
        TH (Sp) & 0.2413 & 0.7037 & 0.9033 & 0.2127 & 0.6559 & 0.8649 \\
        Prec. (Sp) & 0.1818 & 0.6024 & 0.9332 & 0.1813 & 0.6048 & 0.9304 \\
        MB (Sp) & \textbf{0.3800} & 0.5588 & 0.6096 & \textbf{0.3404} & 0.5424 & 0.6124 \\
        EB (Sp) & 0.1974 & 0.6262 & 0.9409 & 0.1828 & 0.5817 & 0.9304 \\
        RS (SS) & 0.1308 & 0.3296 & 0.4883 & 0.1411 & 0.3873 & 0.5127 \\
        VdW (SS) & 0.0966 & 0.2562 & 0.4146 & 0.1236 & 0.2978 & 0.4642 \\
        TH (SS) & 0.1019 & 0.2481 & 0.4072 & 0.1163 & 0.2965 & 0.4764 \\
        Prec. (SS) & 0.1540 & 0.4540 & 0.6774 & 0.1486 & 0.4451 & 0.6748 \\
        MB (SS) & 0.3592 & 0.5463 & 0.6048 & 0.3190 & 0.5322 & 0.5964 \\
        EB (SS) & 0.1856 & 0.5697 & 0.9146 & 0.1718 & 0.5258 & 0.8863 \\
        RE & 0.1382 & 0.4552 & 0.8513 & 0.1412 & 0.4959 & 0.8978 \\
        CM & 0.1859 & 0.4642 & 0.8044 & 0.1815 & 0.4584 & 0.8015
    \end{tabular}
    \caption{The power of several tests under alternative scenarios 5 and 6}
    \label{tab:pow_overcon}
\end{table}

This small power study is not meant to be comprehensive, but is designed so to illustrate some practical cases where various tests based on se-blocks maintain power advantages or comparable power relative to competing methods. One quick take-away from the simulation study is that the linear rank tests with the spiral se-block construction are competitive with the other tests based on se-blocks, the rank-energy test, and the cross-match test in all scenarios, being the most powerful under shifts of a certain magnitude in four of the six scenarios. It is also immediately  clear that the se-block construction impacts the power of the tests. For example, the tests utilizing the stair-step construction appear to have much less power than the tests utilizing the spiral construction in the alternative scenarios concerning an increase in scale. Precisely under these scenarios, the linear rank tests with the spiral construction appear to be very powerful compared to all other tests considered, including the rank-energy and cross-match tests. Finally, due to their generality and simplicity, one may expect the generalized precedence, maximal block, and empty block tests to perform poorly under each scenario. However, this is not the case. For example, the maximal block test is actually the most powerful test under certain shift sizes in the first, fifth, and sixth alternative scenarios. Additionally, the empty block test with the spiral construction consistently outperforms both the rank-energy and cross-match tests in the over-concentration scenarios (Alternative 5 and 6). The conclusion from this brief power study is that multivariate hypothesis tests based on se-blocks maintain competitive power in at least some scenarios.

Given that the multivariate tests based on se-blocks do not rely on asymptotic results, they may also be particularly useful in the sequential hypothesis testing setting, where the experimental or online sets (to be tested against the control) may be quite small, as in the process monitoring setting. Indeed, control charts with estimated parameters may be viewed as a specific instance of the two-sample sequential problem. Before spatial ranks based on measure transportation had been proposed, after considering many existing "distribution-free" multivariate control charts which were developed using spatial signs and ranks, \cite{chen2016distribution} noted that these methods are only asymptotically distribution-free, saying, "Although these nonparametric monitoring methods are able to detect shifts regardless of the underlying distributions, they are not 'distribution-free,' in the sense that the charting procedures are not guaranteed to attain the nominal IC [in-control or null] average run-length ($ARL_0$) without knowing the IC distribution $F_0$." They further noted, "it seems impossible to construct a distribution-free multivariate chart (in a strict and general sense) with a traditional concept of SPC that the control limit is fixed and determined before monitoring." Here, multivariate hypothesis tests based on se-blocks may help to fill this gap. Notably, \cite{holcombe2024distribution} proposed a chart explicitly based on the method proposed in \cite{wald1943extension}. Though not explicitly stated, the proposed control region may be viewed as a collection of se-blocks and is identical to the hyper-rectangular region discussed in the context of the extended precedence test in Section \ref{ext_prec}.

Overall, we note that the proposed tests based on se-blocks have the relative strength in that they are exactly distribution-free, which is a property that many methods based on spatial signs and ranks do not possess. In addition to these favorable properties, tests constructed using se-blocks, particularly tests in the linear-rank family, appear to be competitive in terms of power in comparison with existing nonparametric multivariate tests. Given the ease of use, interpretation, visualization, and access to the sampling distributions for popular test statistics, the multivariate tests based on se-blocks represent a strong alternative to the cross-match test, rank-energy test, and other recently proposed distribution-free tests based on measure transportation theory.

\section{Conclusion}
\label{conc}

In conclusion, this manuscript revisits the concept of se-blocks and demonstrates their utility in deriving many classical nonparametric two-sample hypothesis tests. By unifying several existing tests under the framework of se-blocks, this approach not only simplifies their presentation but also facilitates their extension to high-dimensional settings without altering their null properties. This generalization allows tests to retain exact distribution-free properties and does not require the use of depth functions or spatial signs and ranks, offering a versatile and robust alternative for multivariate nonparametric testing. The generalized testing procedures proposed and considered in this paper are particularly advantageous in settings requiring exact distribution-free procedures and may be useful in diverse areas such as multivariate quality control and life-testing. In revisiting the idea of using se-blocks in the hypothesis testing setting, we also identify several avenues for future research. Though null testing properties are unchanged regardless of dimension, it remains to identify wide classes of alternatives against which these many tests, using various partitioning methods, are consistent and powerful.

Overall, this work aims to stimulate renewed interest in se-blocks, offering a unified perspective that bridges historical methods with modern multivariate testing needs. Future advancements building upon this framework may open new possibilities in both theoretical development and practical application.

\section{Code Availability Statement}

The R scripts and associated files used to implement the proposed methods, partition se-blocks, and generate the figures in this manuscript are available \href{https://osf.io/eypd2/?view_only=bd817ce3a67b44b1bcc76cd538109655}{here}.

\bibliography{refs}

\end{document}